\documentclass[twocolumn]{aastex631}

\usepackage{CJK}
\usepackage{float}
\usepackage{soul}
\usepackage[dvipsnames]{xcolor}

\newcommand{\astrid}{\textsc{ASTRID}}
\newcommand{\Msun}{$\mathrm{M}_{\odot}$}
\newcommand{\circr}{\texttt{Circ}}
\newcommand{\eccr}{\texttt{Ecc}}

\shorttitle{LISA sources in the ASTRID simulation}
\shortauthors{Wang et al.}

\graphicspath{{./}{./Figures/}}

\begin{document}

\begin{CJK*}{UTF8}{bsmi}

\title{Gravitational Waves from Massive 
Black Hole Mergers in ASTRID: Predictions for LISA}

\author[0000-0001-7168-8517]{Bonny Y. Wang (汪玥)}
\affiliation{McWilliams Center for Cosmology and Astrophysics, Department of Physics, Carnegie Mellon University, Pittsburgh, PA 15213, USA}
\affiliation{Entertainment Technology Center, Carnegie Mellon University, Pittsburgh, PA 15213, USA}

\author[0000-0002-8828-8461]{Yihao Zhou (周亦豪)}
\affiliation{McWilliams Center for Cosmology and Astrophysics, Department of Physics, Carnegie Mellon University, Pittsburgh, PA 15213, USA}

\author[0009-0008-0958-8273]{William Chen}
\affiliation{McWilliams Center for Cosmology and Astrophysics, Department of Physics, Carnegie Mellon University, Pittsburgh, PA 15213, USA}

\author[0000-0001-6627-2533]{Nianyi Chen}
\affiliation{School of Natural Sciences, Institute for Advanced Study, Princeton, NJ 08540, USA}

\author[0000-0002-6462-5734]{Tiziana Di Matteo}
\affiliation{McWilliams Center for Cosmology and Astrophysics, Department of Physics, Carnegie Mellon University, Pittsburgh, PA 15213, USA}

\author{Rupert Croft}
\affiliation{McWilliams Center for Cosmology and Astrophysics, Department of Physics, Carnegie Mellon University, Pittsburgh, PA 15213, USA}

\author[0000-0001-5803-5490]{Simeon Bird}
\affiliation{Department of Physics \& Astronomy, University of California, Riverside, 900 University Ave., Riverside, CA 92521, USA}

\author[0000-0001-7899-7195]{Yueying Ni}
\affiliation{Harvard-Smithsonian Center for Astrophysics, Harvard University, 60 Garden Street, Cambridge, MA 02138, USA}

\correspondingauthor{Yihao Zhou}
\email{yihaoz@andrew.cmu.edu}

\begin{abstract}

We use the ASTRID cosmological simulation to forecast massive black hole (MBH) mergers detectable by Laser Interferometer Space Antenna (LISA) down to $z=0$. ASTRID directly models MBH dynamical friction, allowing a realistic tracking of their trajectory. It also incorporates relatively low-mass MBH seeds down to $5\times 10^{4}$ \Msun, providing a more complete picture of LISA MBH mergers. We find that
LISA MBH mergers initially have high eccentricities, peaking around $e_0 = 0.8$ across all redshifts. Accounting for this boosts the event rate from 5.6 yr$^{-1}$ (if circular orbits are assumed) to 10.5 yr $^{-1}$. This enhancement is largely due to additional inspiral sources that will coalesce after LISA's observation, which constitute 46\% of detected events. This underscores the importance of LISA's sensitivity to the early inspiral phase, especially for eccentric binaries that emit gravitational waves across a wider frequency band. 
Most LISA events in ASTRID arise  from $M_{\mathrm{BH}} \sim 10^{5-6}$~\Msun, low-redshift ($z<2$) and low mass-ratio ($q\sim 0.01-0.1)$ mergers.
Accounting for eccentricity broadens the detectable MBH mass range up to $10^9$~\Msun\, and shifts the peak of detectable mergers to a lower redshift $z_{\rm peak} = 0.8$. This implies that the most massive LISA events may also be PTA sources. We predict LISA events to be in various galaxy environments, including many low-mass satellite galaxies. The electromagnetic (EM) counterparts of most LISA sources have active galactic nuclei (AGN) luminosities $L_{\rm bol}> 10^{42}$~erg s $^{-1}$, albeit only $1\%$ with $ > 10^{44}$~erg s $^{-1}$. The brightest AGN are those associated with the rare LISA/PTA events with $M_{\rm BH} > 10^{8}$~\Msun.

\end{abstract}


\section{Introduction} \label{sec:intro}

Massive black holes (MBHs) are ubiquitous in the present Universe, harbored in almost all the massive galactic centers \citep{Kormendy1995, Magorrian1998, Kormendy2013}. 
They play a major role in many aspects of galaxy evolution, and have a tight correlation with their host galaxies \citep{Ferrarese2000, Graham2001, Ferrarese2002, Gutekin2009, McConnell2013, Reines2015, Greene2016, Schutte2019}.
Following the galaxies' merger, their central MBHs migrate toward the center of the remnant galaxy, losing the orbital energy and momentum via dynamical friction (DF) and three-body scattering. After the MBH orbital decay to sub-parsec scales, the MBH binaries emit gravitational waves (GWs) and eventually coalescence \citep{Merritt2013_AGN_dnyamics_book}. The GW signal emitted by the merging MBH offers a unique observational pathway for exploring the formation and evolution of MBHs \citep{Wyithe2003, Sesana2007a, Barausse2012}.

The Laser Interferometer Gravitational Wave Observatory (LIGO) detection 
reveals the population of stellar-mass black holes and validates using GWs to study black hole binaries \citep{Abbott2016}. Recently, multiple Pulsar Timing Arrays (PTAs) have reported the first observation of GWs in the nanohertz frequency band (NANOGrav \citep{Agazie2023_Nanograv_GWB}; CPTA \citep{ Xu2023_CPTA}; PPTA \citep{Reardon2023_PPTA}; EPTA+InPTA \citep{EPTACollaboration2023}), whose proposed sources are primarily expected to be MBH binaries with the mass of $\gtrsim 10^{8}$ \Msun\ \citep{Desvignes2016, Reardon2016, Arzoumanian2018}. Different from LIGO and PTAs, 
the upcoming Laser Interferometer Space Antenna (LISA) mission will target the GWs signals among $10^{-4}-10^{-1}$ Hz, exploring the MBH mergers within the masses range $10^4 - 10^7$ \Msun\ and up to redshifts of $z\sim20$ \citep{AmoraSeoane2017}.

To fully leverage the upcoming LISA detections, it is essential to develop sophisticated theoretical frameworks for merging MBH binaries and to make statistical predictions for the GW event rate. Major efforts have been made in this field in the last decade. Based on semi-analytical galaxy formation models and using black hole seed masses on the order of $10^{4}$ \Msun, some works estimated that the merger rate ranges from $8$ to over $20$ per year \citep{Arun2009, Sesana2011, Klein2016}. 

Some recent predictions are built from large-volume hydrodynamical simulations, which self-consistently follow the co-evolution of the MBH and their host galaxies across cosmic time, capturing nonlinear processes that cannot be described by simple analytical approximations. 
Such simulations can provide a more complete and consistent prediction of the MBH binary population. 
Combining the merging MBHs in the EAGLE simulation suite \citep{Crain2015, Schaye2015} with the phenomenological frequency-domain gravitational waveform model PhenomD \citep{Khan2016}, \citet{Salcido2016} predicted $\sim 2$ detections per year by LISA, and that the initial black hole seed mass could be distinguished through the detected gravitational waveforms. 
\citet{Katz2020} utilized the data from the Illustris cosmological simulation \citep{Vogelsberger2014} and found a detection rate of $\sim 0.5-1$ per year. High-resolution cosmological simulations, albeit with smaller volumes, have also been used as they can better resolve the dwarf galaxies hosting LISA sources \citep[e.g.][]{Tremmel2018, Volonteri2020, Li2022_tng50, Li2023}.

In this work, we study the MBH mergers that are detectable by LISA in the large-volume cosmological simulation \astrid\ \citep{Bird2022, Ni2022, Ni2024}, which recently arrived at $z=0$. 
\astrid\ uses a novel power-law seeding with a range of MBH seed masses and so includes relatively low-mass MBH ($\sim 5\times 10^{4}$ \Msun). 
More importantly, a subgrid model developed based on \citet{Tremmel2015} and \citet{ Chen2022_DF} is incorporated in \astrid\ to track the MBH dynamical friction on the fly. 
This enables us to follow the MBH dynamics down to the resolution limit ($\sim$ 1kpc), and to measure the realistic orbital evolution.
These advantages put \astrid\ in a unique position to be used to study MBH merging
population.
\citet{Chen2022} and \citet{DeGraf2024} investigate the high-redshift ($z > 2$) MBH mergers population and the host galaxies detectable by LISA. With \astrid\ reaching down to $z = 0$, we now have the opportunity to examine sources at lower redshifts, to which LISA is most sensitive \citep{Salcido2016, Izquierdo-Villalba2024}.

This paper is organized as follows: in Section \ref{sec:methods} we introduce the \astrid\ simulation, and describe the methods used to calculate the GW signals. In Section \ref{sec:results} we present the MBH population in \astrid, and make predictions for the LISA detection rate. 
In Section~\ref{sec:hostgal}, we analyze the properties of the host galaxies and the MBH population of the LISA sources. 
We then discuss the implications and caveats of our results in Section~\ref{sec:discussion} and conclude in Section~\ref{sec:conclusion}.

\section{Methods}
\label{sec:methods}

\begin{figure*}
    \centering
    \includegraphics[width=1 \linewidth]{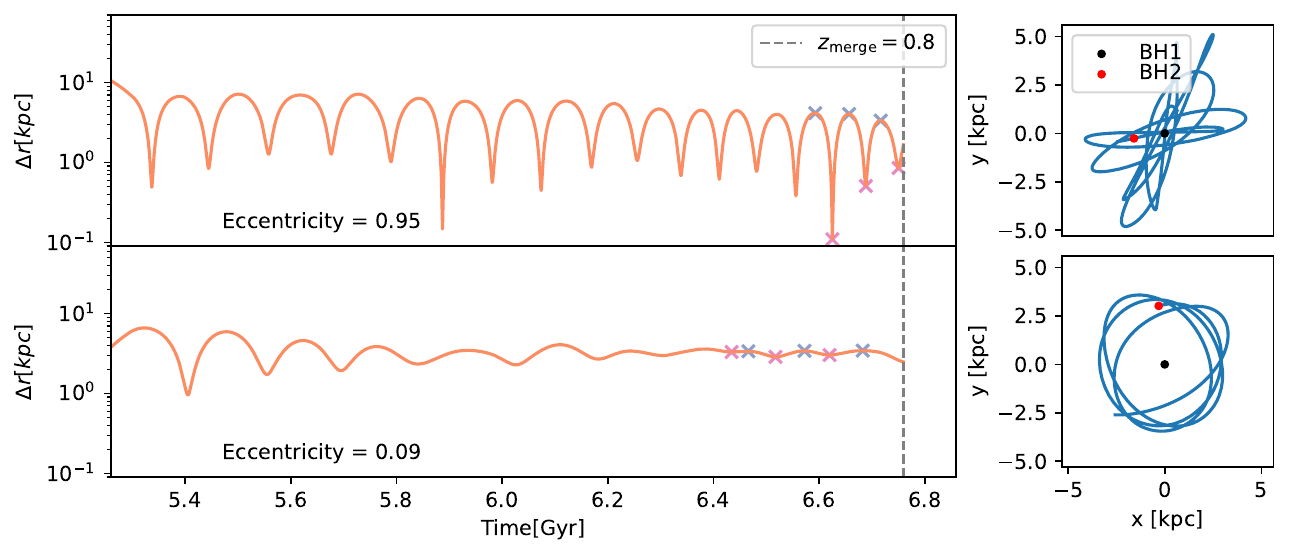}
    \caption{Examples of MBH mergers with a circular orbit ($e = 0.09$; upper panel) and a highly eccentric orbit ($e = 0.95$; lower panel) in \astrid\ at $z=0.8$. The plots on the left display the separation $\Delta r$ between the two merging black holes as a function of time, with pink and blue crosses marking the last three pairs of $r_{\text{apo}}$ and $r_{\text{peri}}$ used to calculate the eccentricity values. The plots on the right show the corresponding orbital trajectories of the secondary black hole relative to the primary black hole's position. The black and red dots mark the locations of the primary black hole (BH1) and the secondary black hole (BH2) before the merger.}
    \label{fig:orbit}
\end{figure*}

\subsection{The ASTRID Simulation}
\astrid\ \citep{Bird2022,Ni2022,Ni2024, Zhou2025_redshift0} is a large-volume cosmological hydrodynamical simulation performed using the Smoothed Particle Hydrodynamics (SPH) code \textsc{MP-Gadget} \citep{Feng2018}. The simulation evolves a cube of 250~$h^{-1}$cMpc per side with $2\times5500^3$ initial tracer particles comprising dark matter and baryons. The cosmological parameters used are from  the \cite{Planck} cosmology, with $h=0.6774$, $\Omega_{0}=0.3089$, $\Omega_{\mathrm\Lambda}=0.6911$, $\Omega_{\mathrm{b}} = 0.0486$, $\sigma_{\mathrm{8}}=0.816$, and $n_{\mathrm{s}}=0.9667$. \astrid\ has a mass resolution of $M_{\rm DM} = 6.7 \times 10^6$~$h^{-1}M_\odot$  and $M_{\rm gas} = 1.3 \times 10^6$~$h^{-1}M_{\odot}$  in the initial conditions. 
The gravitational softening length is $e_{\rm g} = 1.5 h^{-1}{\rm ckpc}$ for both DM and gas particles. 
The initial conditions are set at $z=99$ and it has been evolved to $z=0$. 

\astrid\ includes a full-physics sub-grid treatment for modeling galaxy formation, MBHs and their associated supernova AGN feedback, as well as inhomogeneous hydrogen and helium reionization. 
In the following we briefly list the sub-grid models relevant to MBH evolution. We refer readers to \cite{Bird2022} and \cite{Ni2022} for more details of the implemented models, and to \cite{Chen2022} for a full description of the high-redshift merger catalog.

MBHs in \astrid\ are seeded in halos with $M_{\rm halo,FOF} > 5 \times 10^9~h^{-1}\rm{M_\odot}$ and $M_{\rm *,FOF} > 2 \times 10^6$~$h^{-1}\rm{M_\odot}$, with seed masses stochastically drawn from the range $3\times10^{4}~h^{-1}\rm{M_\odot}$ to $3\times10^{5} ~h^{-1}\rm{M_\odot}$.
The gas accretion rate $\dot{M}_{\mathrm{BH}}$ is estimated using the Bondi-Hoyle formalism \citep{BondiHoyle1944}. 
Super-Eddington accretion is allowed with an upper limit of twice the Eddington accretion rate. 
Each MBH radiates with a bolometric luminosity $L_{\rm bol}$ proportional to the accretion rate $\dot{M}_\mathrm{BH}$, computed using a mass-to-energy conversion efficiency $\eta=0.1$ \citep{Shakura1973}. Of this, 5\% of the radiated energy is thermally coupled to the gas within twice the radius of the SPH smoothing kernel of the black hole particle. 
A subgrid model is applied to estimate MBH dynamical friction following \cite{Tremmel2015} and \cite{Chen2022}.
Compared to the more common implementation that directly repositions BHs to the local potential minimum \citep{Weinberger2017_tng}, this model gives well-defined MBH trajectories and velocities.
Two MBHs merge if their separation is within two times the gravitational softening length $2\epsilon_g$, once their kinetic energy is dissipated by DF and they are gravitationally bound to the local gravitational potential.

\subsection{Shape Measurement}
\label{sec:shape}

In the \astrid\ simulation, the trajectories of MBHs are output with high time resolution (every single simulation timestep).  Having this fully sampled data leading
up to each merger event enables accurate analysis of their orbital evolution. 
To measure the orbital eccentricity $e_0$, we employ the shape measurement techniques described in \citet{Chen2022}, which we now briefly summarize.
We use the subscript of $0$ to indicate the orbital eccentricity is used as the initial condition in the evolution during the GW regime, which will be changed due to the GW circularization \citep{Peters1963}. 
Taking the distance between the two merging black holes 
$\Delta r$, we define the periapsis $r_{\rm peri}$ to be its local minimum, and the apoapsis $r_{\rm apo}$ to be its local maximum. The eccentricity, $e_0$, is then given by:
\begin{equation}
\label{eq:ecc_cal}
    e_0 = \frac{r_{\rm apo} - r_{\rm peri}}{r_{\rm apo} + r_{\rm peri}}.
\end{equation}
For each merger, we compute $e_0$ from the three sets of  $r_{\mathrm{apo}}$ and $r_{\mathrm{peri}}$ values that immediately precede the merger. We then select the maximum $e_0$ from these three values to be the final measurement of $e_0$ for each merger. 
This procedure minimizes the impact of falsely identified local extrema, which can produce unrealistically small $e_0$.
Without this choice, we would not correctly quantify the effect of incorporating eccentricity when compared to circular orbits. 
Figure \ref{fig:orbit} illustrates an example of 
a highly eccentric merging orbit (\(e_0 \approx 1\), upper panel) and a circularized orbit (\(e_0 \approx 0\), lower panel) at redshift 0.8. The left panels display \(\Delta r\) as a function of time, where
the pink and blue crosses indicate the identified \(r_{\rm peri}\) and \(r_{\rm apo}\) values, which are used for the eccentricity calculation. 
The right panels show the corresponding orbits projected onto a plane perpendicular to the average relative velocity between the MBHs.

\subsection{Gravitational Wave Signal Calculation}
\label{sec:GWCalc}

\begin{figure*}
    \centering
    \includegraphics[width = 1\textwidth]{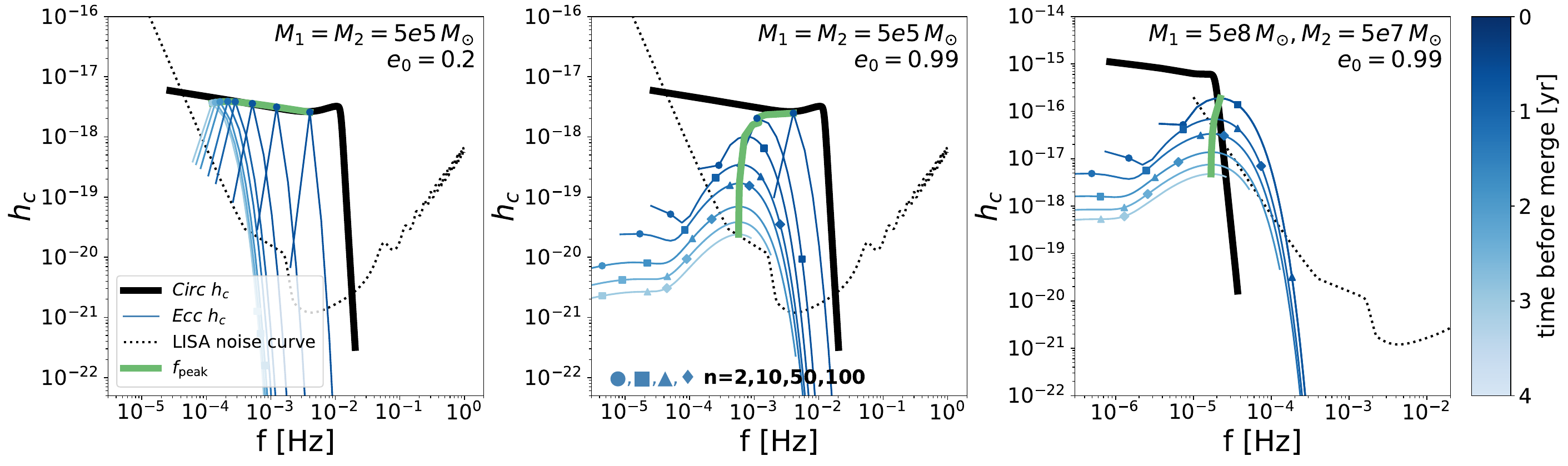}
    \caption{Waveforms for three binaries of different mass and orbital eccentricity $e_0$ during the 4 years before the coalescence. The strain is calculated for $z=1$, and the mass and the initial $e_0$ are labeled on the upper right corner. 
    In each panel, the black dotted line is the LISA instrument sensitivity curve with the galactic binary background. 
    The thick black curve shows the strain assuming initial $e_0=0$. The thin blue curves track the strain at different evolution times. The lines of constant time are shaded from light to dark for early to late evolution. 
    On each blue curve, the circle, square, triangle, and diamond markers represent the evolution of individual harmonics: n=2, 10, 50, and 100, respectively.
    The thick green curve plots the frequency having the highest strain $f_{\mathrm{peak}}$ for the eccentric waveforms during the evolution.
    }\label{fig:ecc_waveform}
\end{figure*}

We use the \texttt{gwsnrcalc}\footnote{\url{https://pypi.org/project/gwsnrcalc/}} package \citep[][]{Katz2019} to calculate the GW emitted by MBH mergers. We generate two sets of GW signals, one assuming circular orbits (labeled as \texttt{Circ}), and our main prediction, which incorporates the orbital eccentricity inferred (and evolved) from the BH orbits in ASTRID, as described in Section~\ref{sec:shape}. We label the predictions using the eccentricity information as \texttt{Ecc}. 

The  \texttt{Circ} GW signals are from the PhenomD phenomenological model \citep[][]{Husa2016,Khan2016}. This model requires input parameters such as the binary masses, the merger redshift, and the dimensionless spin of the binary components. 
The spin parameter \(a\) is a measure of the alignment between the spin angular momentum and the orbital angular momentum. Since $a$ is not available in our simulation, we adopt a fixed spin value of \(a_1 = a_2 = 0.8\) for all binaries, as proposed in \citet{Katz2020} and used in many prior researches \citep[e.g.,][]{Miller2007,Reynolds2013,Chen2022, DeGraf2024}. 
In PhenomD, the characteristic strain for a circular source, \(h_s\), which accounts for the time the binary spends in each frequency ($f$) bin, is defined as:
\begin{equation}
   \label{eq:hc}
       h_s(f) = 4f^2 |\tilde{h}(f)|^2,
\end{equation}
where \(\tilde{h}(f)\) represents the Fourier transform of the time-domain signal \citep{Finn2000,Moore2015}.

For \texttt{Ecc}, 
gravitational wave strain is influenced by the orbital eccentricity. The $h_{\mathrm{s}}$ from an individual eccentric source can be related to that of a circular source as \citep[see ][]{Amaro-Seoane2010,Kelley2017b, Chen2022}:
\begin{equation}
\label{eq:ecc_hf}
    h_s^2(f_{\rm r}) = \sum_{n=1}^{\infty} \left(\frac{2}{n}\right)^2  h_{\rm r,circ}^2 (f_{\rm h}) g(n,e) \big|_{f_{\rm h} = f_{\rm r}/n},
\end{equation}
where \(h_{\rm r,circ}\) is the characteristic strain for a circular source, as given in equation~\ref{eq:hc}. \(g(n,e)\) represents the GW frequency distribution function, defined by equation~20 in \citet{Peters1963}.

PhenomD describes the waveforms in the inspiral-merger-ringdown (IMR) phase in the $l=m=2$ quasinormal mode. 
For an eccentric binary merger, the inspiral waveforms are sufficiently accurate, but waveform models including the contributions from the merger and ringdown stages are less developed. 
Although recent efforts have been made to construct  IMR waveforms with eccentricity \citep{PhysRevD.98.044015,PhysRevD.103.124011,Chattaraj2022}, 
some important physical effects such as spin or higher order modes are not included, and 
there is still a lack of a clear and simple model to be implemented. Hence, to make a better comparison between the \circr\ and \eccr, in this work we only consider the waveforms from the inspiral phase while not including the emission during the phase of merger and ringdown. 

To better assess the detectability of MBH binaries, we calculate the signal-to-noise ratio (SNR) for each merger. The SNR is determined by integrating the $h_{\mathrm{s}}$ above the LISA sensitivity curve over the frequency domain. 
Specifically, we compute the sky-averaged, orientation-averaged, and polarization-averaged SNR using the following expression:
\begin{equation}
\label{eq:snr}
\langle {\rm SNR} \rangle^2 = \frac{16}{5}\int_{f_{\mathrm{start}}}^{f_{\mathrm{end}}} \frac{h_s^2}{h_N^2} f^{-1} df,
\end{equation}
where \(f_{\rm start} = f(t_{\rm start})\) and \(f_{\rm end} = f(t_{\rm end})\), with \(t_{\rm start}\) and \(t_{\rm end}\) denoting the beginning and end times of the observed signal. $h_\mathrm{s}$ is calculated based on Eq.~\ref{eq:hc} and Eq.~\ref{eq:ecc_hf} for \texttt{Circ} and \texttt{Ecc}, respectively.
$h_{\mathrm{N}}$ is the strain of the LISA noise curve, which is from the LISA Mission Proposal \citep{Amaro-Seoane2017}.
In this work, we include the effect of galactic background noise in addition to the LISA instrumental noise, for which we use the analytical approximation from \citet{Hiscock2000}. We classify mergers with \(\mathrm{SNR} > 10\) as sources detectable by LISA.

In Fig.~\ref{fig:ecc_waveform}, we show the waveforms for three binaries with different masses and orbital eccentricities $e_0$ during the four years before coalescence. 
The strain is calculated for $z=1$, and the mass and the initial $e_0$ are labeled in the upper right corner. 
In each panel, the black dotted line is the LISA instrument sensitivity curve with the galactic binary background. 
The thick black curve shows the strain assuming an initial $e=0$. The thin blue curves track the strain at different evolution times. The lines of constant time are shaded from light to dark for early to late evolution. 
On each blue curve, the circle, square, triangle, and diamond markers represent the evolution of individual harmonics: n=2, 10, 50, and 100, respectively.
We do not mark all the harmonic modes for visual clarity.
The thick green curve plots the frequency having the highest strain $f_{\mathrm{peak}}$ for the eccentric waveforms during the evolution. 
For the binaries with the circular orbit, the strain comes from the individual harmonic mode $n=2$; while for the eccentric binaries, the GW signal is contributed by all $n\geq1$ harmonic modes, which produces a wide frequency range. If the circularization causes the eccentricity close to zero in the end (the left and middle panels), we find that the behavior of the \texttt{Ecc} is very similar to that of the \texttt{Circ} as the thick green curve coincides with the $n = 2$ modes, indicated by the blue circles. 
Since the total SNR is typically dominated by the emission in the short phase before coalescence, these fully circularized sources have similar detectability in both the \eccr\ and \circr.
While for the massive mergers shown in the right panel, incorporating eccentricity boots the detectability. This is because the high emission modes produce emissions closer to the high-sensitivity band, contributing significantly to the total SNR.

The signal from high harmonic modes ($n\geq2$) is important for the high-eccentricity sources: for binaries with $e=0.99$, the dominant mode at 4 years before the coalescence is larger than n=100 (see the low-frequency end of the $f_{\mathrm{peak}}$ curves). To ensure we have sufficient harmonic modes for the high-eccentricity sources while keeping the computational cost reasonable, we truncate the summation in Eq.~\ref{eq:ecc_hf} at $n_{\mathrm{max}}$ for the sources in \eccr, and vary the value of $n_{\mathrm{max}}$ with the initial eccentricity $e_0$.

To determine $n_{\mathrm{max}}$, we first search for the dominant harmonic mode that has maximum strain based on equation 20 of \citet{Peters1963}. This $n_{\rm dom}$ is independent of the mass or redshift for specific sources, while only determined by $e_0$ \citep{Hamers2021,Wagg2022}.

We then set $n_{\mathrm{max}} = 2 \times n_{\mathrm{dom}}$ with a minimum value of $n=50$, i.e., 
$n_{\mathrm{max}}=\max \left(2 \times n_{\mathrm{dom}},\ 50\right)$. 

\subsection{Merger Rate Prediction}
\label{sec:MergerRatePredict}

To estimate the LISA detection rate, we follow \citet{Katz2020} and implement the Monte Carlo analysis technique.
This enables us to generate multiple realizations of the merger population detected by LISA. In the following we briefly summarize this method and refer readers to \citet{Katz2020} for more details.

Based on the MBH population in \astrid, the MBH merger rate is defined as the number of merging MBH systems observed per year per unit redshift, which can be estimated from:
\begin{equation}
\label{equ:MergerRate}
    \frac{{\rm d}^{2}N(z)}{{\rm d}z\,{\rm d}t} 
    =\frac{\Delta N}{\Delta z\,V_{\mathrm{sim}}} \frac{{\rm d}z}{{\rm d}t} \frac{{\rm d}V_c(z)}{{\rm d}z} \frac{1}{1+z},
\end{equation}
where $dV_{\mathrm{c}}(z)$ is the comoving volume element of the Universe at a given redshift, and $V_{\mathrm{sim}}$ is the comoving volume of \astrid\ simulations. 
$\Delta N(z)$ is the number of mergers in a specific redshift bin, whose width is $\Delta z$.
For each MBH merger in \astrid, its contribution to the merger population observed within a $t_{\mathrm{dur}}$ time window is 
\begin{equation}
    \lambda_i
    =t_{\mathrm{dur}}\,\frac{{\rm d}V_c(z_i)}{V_{\mathrm{sim}}} \frac{1}{1+z_i},
\end{equation}
where $z_i$ is the merger redshift of the MBH binary $i$. 
The expected total source number for one realization of the observed MBH merger population is then given by 
\begin{equation}
\left<N_{t_{\mathrm{dur}}} \right>
    = \frac{t_{\mathrm{dur}}}{V_{\mathrm{sim}}}\sum_{i}\lambda_i\frac{dV_{c}(z_i)}{1+z_i},
\end{equation}
where $\sum_{i}$ denotes the sum over all the mergers in \astrid. 
To generate multiple realizations, we draw values from the Poisson distribution centered at $\lambda_{i}$: $\mathcal{P}(\lambda_i)$ to determine whether the binary $i$ is included in specific realizations or not.
Hence, the  number of sources for one realization is 
\begin{equation}
    N_{t_{\mathrm{dur}}} 
    = \frac{t_{\mathrm{dur}}}{V_{\mathrm{sim}}}\sum_{i}\mathcal{P}(\lambda_i)\frac{dV_{c}(z_i)}{1+z_i}.
\end{equation}
Specifically, if $\mathcal{P}(\lambda_i)\geq 2$, we use the same binary parameters (e.g., the MBH masses, merger redshift, and eccentricity) for the duplicated mergers to calculate the expected GW emission. 
In this work, we generate 10,000 realizations to predict the
LISA detection rate.

An advantage of implementing this Monte Carlo analysis is that we can include the GW signals emitted by the inspiral sources, which coalesce after the LISA mission ends. To do this, we sample the MBH populations merging within $t_{\mathrm{dur}}$, and the first $t_{\mathrm{obs}}$ phase $(t_{\mathrm{dur}}>t_{\rm obs})$ corresponds to the observation time. 
The binaries which take less time than $t_{\mathrm{obs}}$ to merge are marked as `merger sources', which are observed to coalesce.
The binaries which merge after  $t_{\mathrm{obs}}$ are labeled as `inspiral sources' as 
they keep inspiraling during the whole observation phase.  
For the LISA mission, $t_{\mathrm{obs}}=4$ years. Following \citet{Katz2020}, we set $t_{\mathrm{dur}}=100$ years \footnote{We test different choices of $t_{\mathrm{dur}}$, and confirm that our prediction results converge with  $t_{\mathrm{dur}}\geq 100$ yr.}.
As the Universe evolves negligibly over $t_{\mathrm{dur}}$, 
we can assume
that the merger times of the $N_{\mathrm{t_{\rm dur}}}$ sources are uniformly distributed over the interval of length $t_{\mathrm{dur}}$.
We therefore draw the observation start $t_{\mathrm{start}}$ randomly from between $0$ and $100$ years to decide which 4-year phase for an individual merger is observed during $t_{\mathrm{dur}}$ beyond coalescence, 
For the observation end time $t_{\mathrm{end}}$, $t_{\mathrm{end}}=t_{\mathrm{start}}-t_{\mathrm{obs}}$ if $t_{\mathrm{start}} > t_{\mathrm{obs}}$ (inspiral sources); and $t_{\rm end}=0$ if $t_{\mathrm{start}} \leq t_{\mathrm{obs}}$ (merger sources). GW signals from $t_{\rm start}$ to $t_{\rm end}$ before the coalescence are used to calculate the SNR for each merger. 
We use the $e_0$ measured by equation~\ref{eq:ecc_cal} as the initial eccentricity at the beginning of LISA observation ($t_{\rm start}$), and it will be circularized with the GW emission according to equation 5.13 in \citet{Peters1964}.
We do not include the effect of circularization before the observation starts as the eccentricity decay rate, which depends on $a^{-4}$ ($a$ is the length of semi major axis of the binary orbit), is typically small during the early inspiral phase.

\begin{figure}
    \centering
    \includegraphics[width=0.99\linewidth]{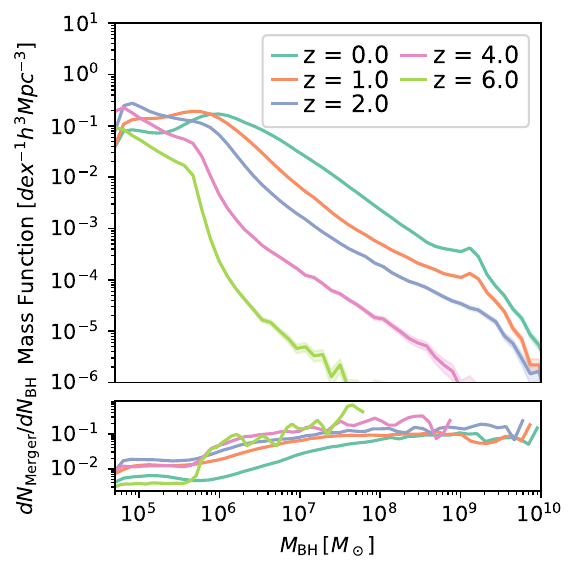}
    \caption{The MBH population in \astrid\ at different redshifts. $Upper:$ The black hole mass function in terms of comoving number density (\(\mathrm{dex}^{-1} h^3 \mathrm{Mpc}^{-3}\)).
    $Lower:$ The ratio between the number of the MBH involved in mergers and the total MBH.}
    \label{fig:BMF}
\end{figure}

\begin{figure*}[!htb]
    \centering
    \includegraphics[width=1\linewidth]{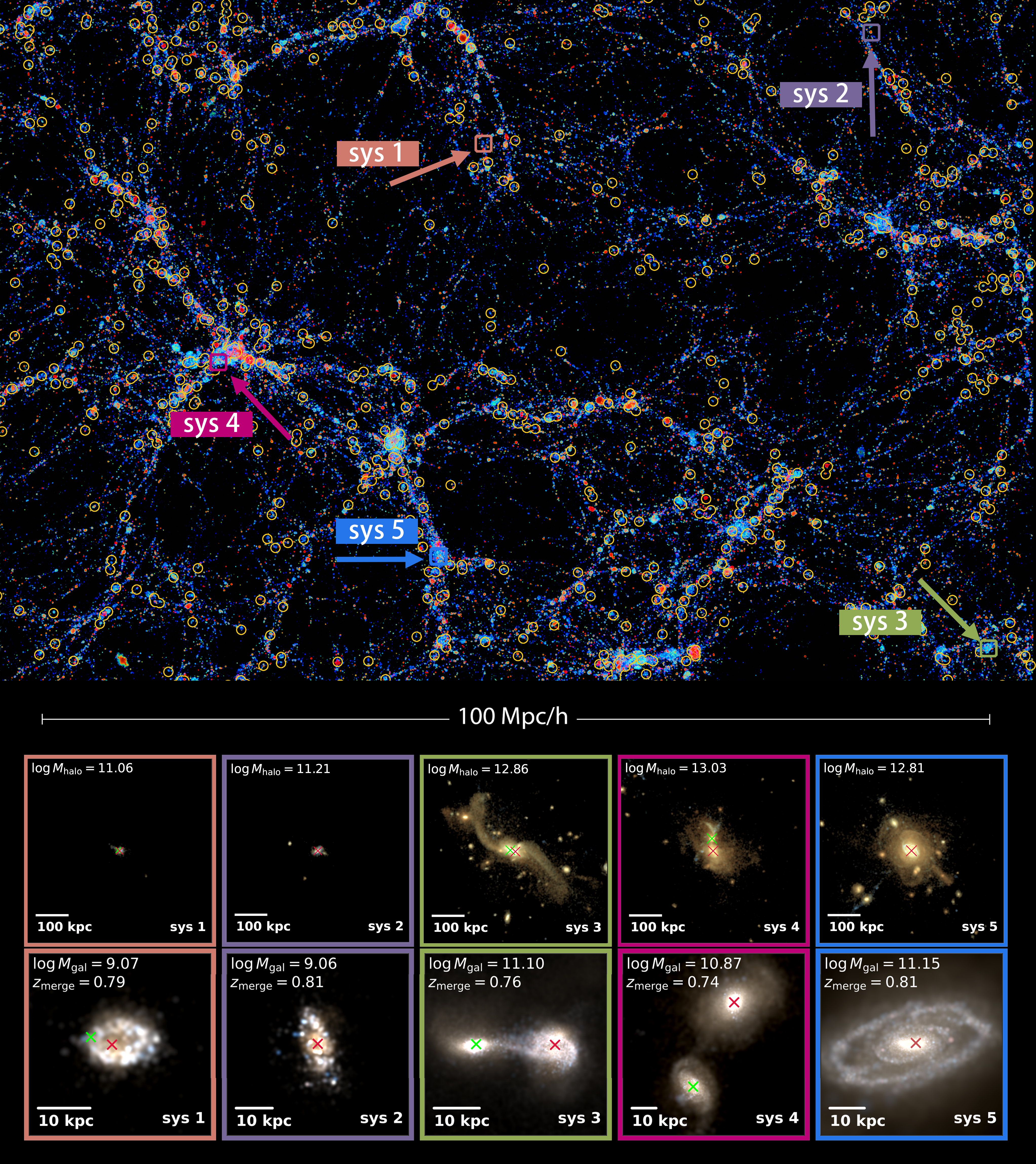}
    \caption{Illustration of \astrid\ at $z=0.8$.
    The upper panel shows the large-scale structure in a slice of $100 \times 60 \times 10\ \mathrm{Mpc}^{3}\,h^{-3}$,  with gas temperature represented by color: hotter regions appear in red, while cooler regions are shown in blue. Yellow circles mark the locations of detectable mergers (i.e., with SNR $>10$) occurring within the 670 Myr around $z=0.8$. 
    Five merging systems are selected as examples, whose positions are marked in the plot. 
    The middle panels show the stellar density field within $600\ \mathrm{kpc}/h$ around the merging MBH binaries, color-coded by the stellar age, where blue regions represent young stars. 
    The red crosses show the position of the primary MBH (or the remnant MBH), the green crosses show the secondary MBH. 
    The bottom panels show the host galaxy of the merging MBHs, with the RGB channels representing the flux in the rest-frame $grz$ color bands.}
    
    \label{fig:LSS}
\end{figure*}

\begin{figure*}[!htb]
    \includegraphics[width = 1\linewidth]{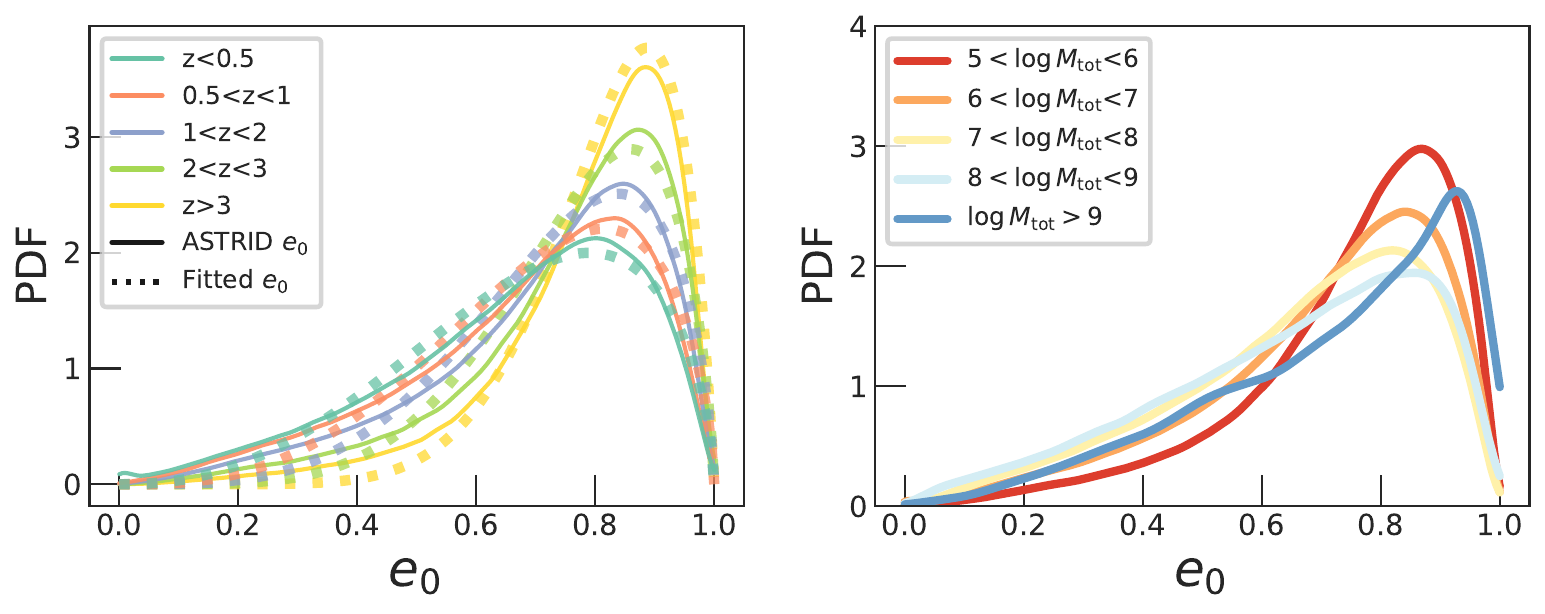}
    \caption{Probability density function (PDF) of merger orbit eccentricities $e_0$ measured based on Eq.~\ref{eq:ecc_cal}. 
    \textit{Left:} PDF for mergers at different redshifts.
    Solid curves represent the distribution from \astrid\ merger population, and dotted curves show the fitted $\beta$ distribution (Equ.~\ref{equ:fitted_ecc_pdf}) for the redshifts with the corresponding color.
    \textit{Right:} PDF for mergers with different total MBH masses.
    }
    \label{fig:ElipDis_fit}
\end{figure*}

\begin{figure*}[!htb]
    \centering
    \includegraphics[width=0.99\linewidth]{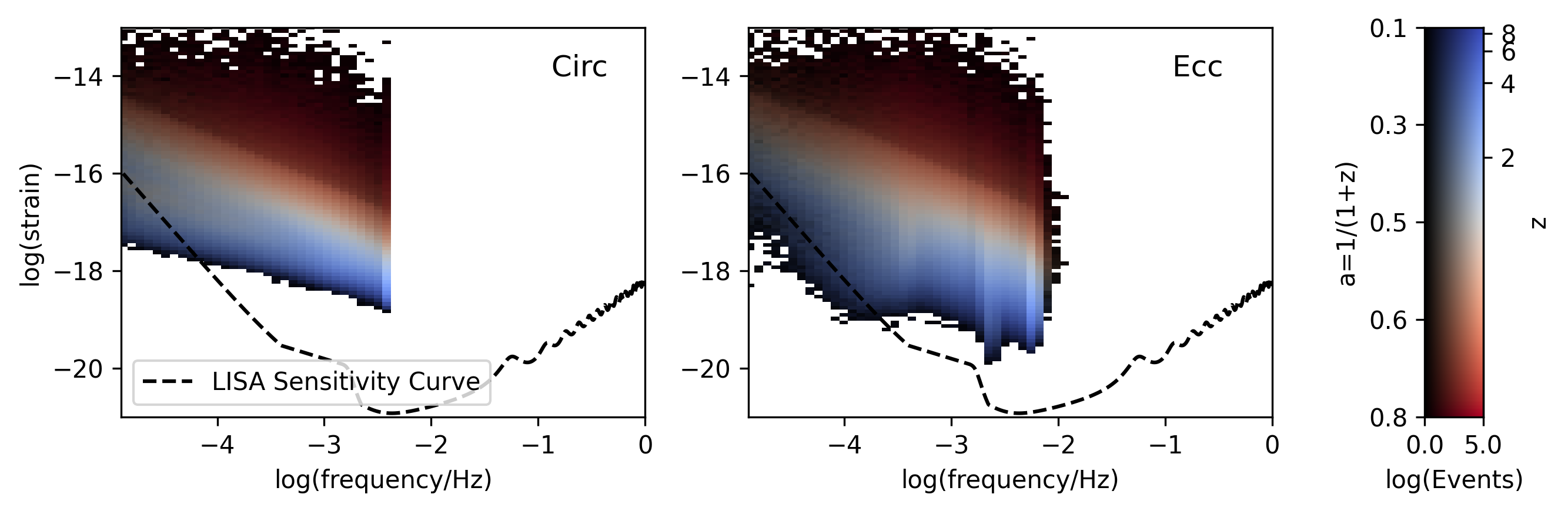}
    \caption{Strain and frequency distribution for the mergers in ASTRID. The left frame is for \circr\ and the right panel is for \eccr. 
    For each MBH merger, the plotted frequency and strain correspond to the evolution time point and harmonic mode that makes the largest contribution to SNR within 4 years before the coalescence.
    For the \texttt{Circ}, the harmonic mode is always $n=2$.  
    We color the distribution based on the redshift and the number of events, as shown by the right colorbar: the high(low) redshifts are represented by the blue(red) colors, and the darkness corresponds to the regions with very few events. 
    The redshift is averaged over the merger population in each pixel. 
    }
    \label{fig:strain_freq}
\end{figure*}

\begin{figure*}[!htb]
    \centering

    \includegraphics[width=1\linewidth]{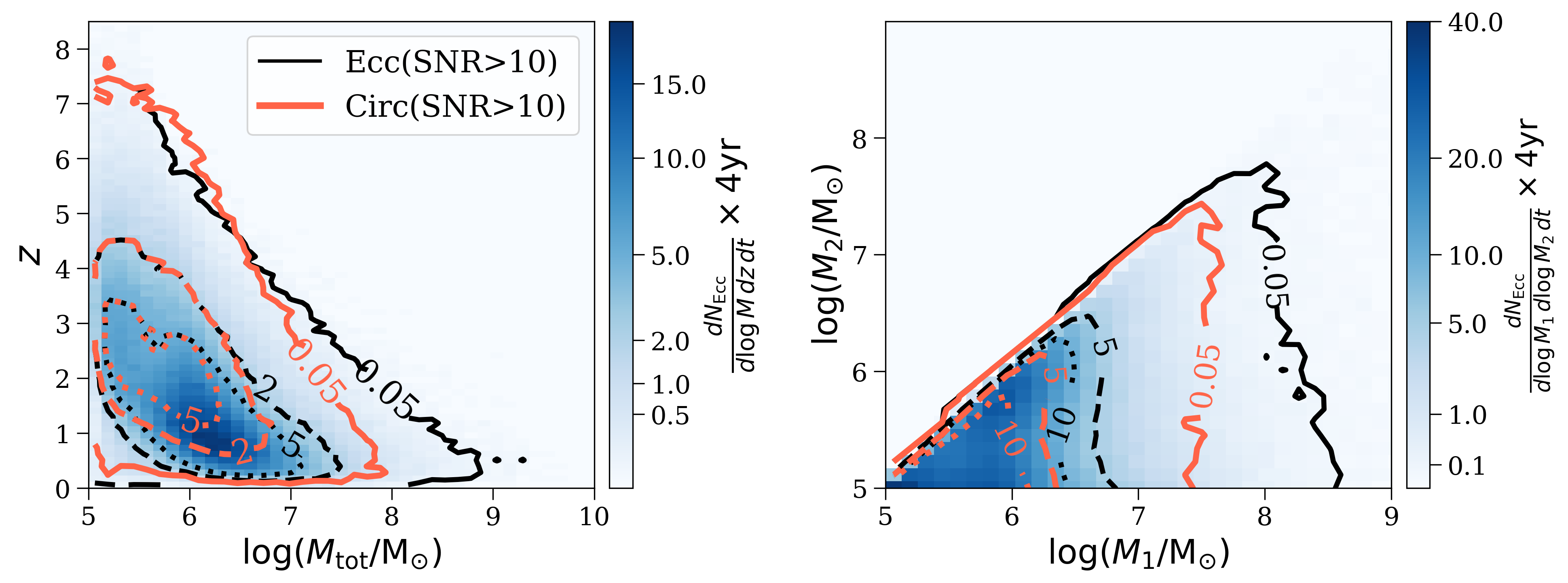}
    \caption{Parameter space of MBH mass and redshift explored by LISA.  
    The left panel plots the two-dimensional distribution of the redshift and the total mass of merging MBHs $M_{\mathrm{tot}}$. The right panel shows the distribution of the mass of the primary MBH $M_1$ and the secondary MBH $M_2$. 
    The underlying blue color gradient represents the differential number of detected mergers during the LISA 4-year observation for \eccr: $dN/(d\log M\, dz\,dt)
\times \mathrm{4yr}$ for the left panel and $dN/(d\log M_1\, d\log M_2\,dt)
\times \mathrm{4yr}$ for the right panel. 
    The black and red contours correspond to the detected MBH merger population for \eccr\ and \circr, respectively. 
    }
    \label{fig:SNR_dist}
\end{figure*}

\section{Results}
\label{sec:results}

In this section, we present the merging MBH populations in the \astrid\ simulation, and make predictions for the sources to be detected by the LISA mission.

\subsection{MBH Population in ASTRID}

In Figure~\ref{fig:LSS}, we present an illustration of LISA sources in a slice of
$100 \times 60 \times 10\ \mathrm{Mpc}^{3}\,h^{-3}$
extracted from \astrid\ at $z=0.8$. 
The upper panel illustrates the gas density field color-coded by the temperature, with warmer regions corresponding to higher temperatures.\footnote{The visualization was created using \texttt{Blender} software; see the companion video at \url{https://astrid.psc.edu/gallery/}}
Yellow circles mark the locations of the mergers occurring within a 670 Myr period centered on $z=0.8$.
We select five merging systems as examples, which cover the whole mass range ($10^{5}\sim 10^{8}\ \mathrm{M}_{\odot}$) for the LISA sources, and mark their position with arrows. 
The middle panels show the stellar density field of the host halo for the five systems, color-coded by stellar age, where blue regions represent young stars. 
The red crosses show the position of the primary MBH (or the remnant MBH), and the green crosses show the secondary MBH. 
The bottom panels show their host galaxy, with the RGB channels representing the flux in the rest-frame $grz$ color bands.

In the upper panel of Figure~\ref{fig:BMF}, we present the black hole mass function (BHMF) in \astrid\ at different redshifts ($z = 0.0, 1.0, 2.0, 4.0, 6.0$). 
The BHMF is defined as the number density of black holes per unit comoving volume as a function of their mass, expressed as \(\Phi(M_{\mathrm{BH}}, z) = \frac{dN}{d({\log M}_{\mathrm{BH}}) dV_c}\), where \(dN\) is the number of black holes within a mass bin, $d({\log M}_{\mathrm{BH}})$ is the width of that mass bin, and \(dV_c\) is the comoving volume element at $z$. 
It can be seen that at lower redshifts, there is an increasing abundance of MBHs. The peak of the distribution shifts from the seed mass range ($10^{5}\,\mathrm{M}_{\odot}$) at $z=6.0$ to approximately \(10^6 M_\odot\) at $z<1.0$. 
The lower panel shows the ratio of the number of merged black holes to the total black hole population. 
At the lower mass end, the MBHs merge less efficiently at lower redshifts after $z=2$. 
This can be explained by the lower galaxy merger rate at low-redshift \citep{Rodriguez-Gomez2015}.

In Figure~\ref{fig:ElipDis_fit} we present the merging orbit eccentricity $e_0$ distributions, which are estimated based on the method described in Section~\ref{sec:shape}. 
The left panel shows the probability density function (PDF) for $e_0$ at different redshifts. Across the whole redshift range, $e_0$ peaks at high values: $e_0 \gtrsim 0.8$, and it evolves to a smaller value at lower redshift: from $0.88$ at $z>3$ to $0.80$ at $z<0.5$, indicating a larger fraction of binaries at low redshift merge with circular orbits. 
A potential explanation for this evolution is that binaries with eccentric orbits have shorter DF timescales \citep{Jiang2008, Boylan-Kolchin2008}, and they will merge shortly after the binaries are formed in the simulation, while many circular binaries keep evolving towards lower redshift. To give a better description of the $e_0$ distribution, we fit the PDF shown in Fig.~\ref{fig:ElipDis_fit} using a $\beta$ distribution:
\begin{equation}
\label{equ:fitted_ecc_pdf}
    f\left(e\right) = \mathcal{N} e^{\alpha-1} \left(1-e\right)^{\beta-1}, \ \ 0\leq z\leq 1.
\end{equation}
The normalization factor is $\mathcal{N}=1/\mathrm{B}(\alpha, \beta)$, where $\mathrm{B}\left(\alpha,\beta \right)$ is the $\beta$ function. The two shape parameters $\alpha$ and $\beta$  depend linearly on redshift $z$: $\alpha=1.08\,(1+z)+2.08$, $\beta=0.07\,(1+z)+1.56$. The fitted distributions are plotted in Figure~\ref{fig:ElipDis_fit} with dotted curves. 
The right panel shows the PDF within different total MBH masses. Although the evolution of $e_0$ over $M_{\mathrm{BH}}$ is less obvious compared to that over $z$, we note that the most massive mergers $M_{\mathrm{tot}}>10^{8}$ \Msun\ typically have higher $e_0$.

\subsection{LISA Predictions}
In this subsection, we investigate the MBH mergers in the \texttt{Circ}/\texttt{Ecc} that are detectable by LISA (i.e., with SNR $>10$), and make predictions for the LISA detection rate.

\begin{figure*}[!htb]
    \centering
    \includegraphics[width=1\textwidth]{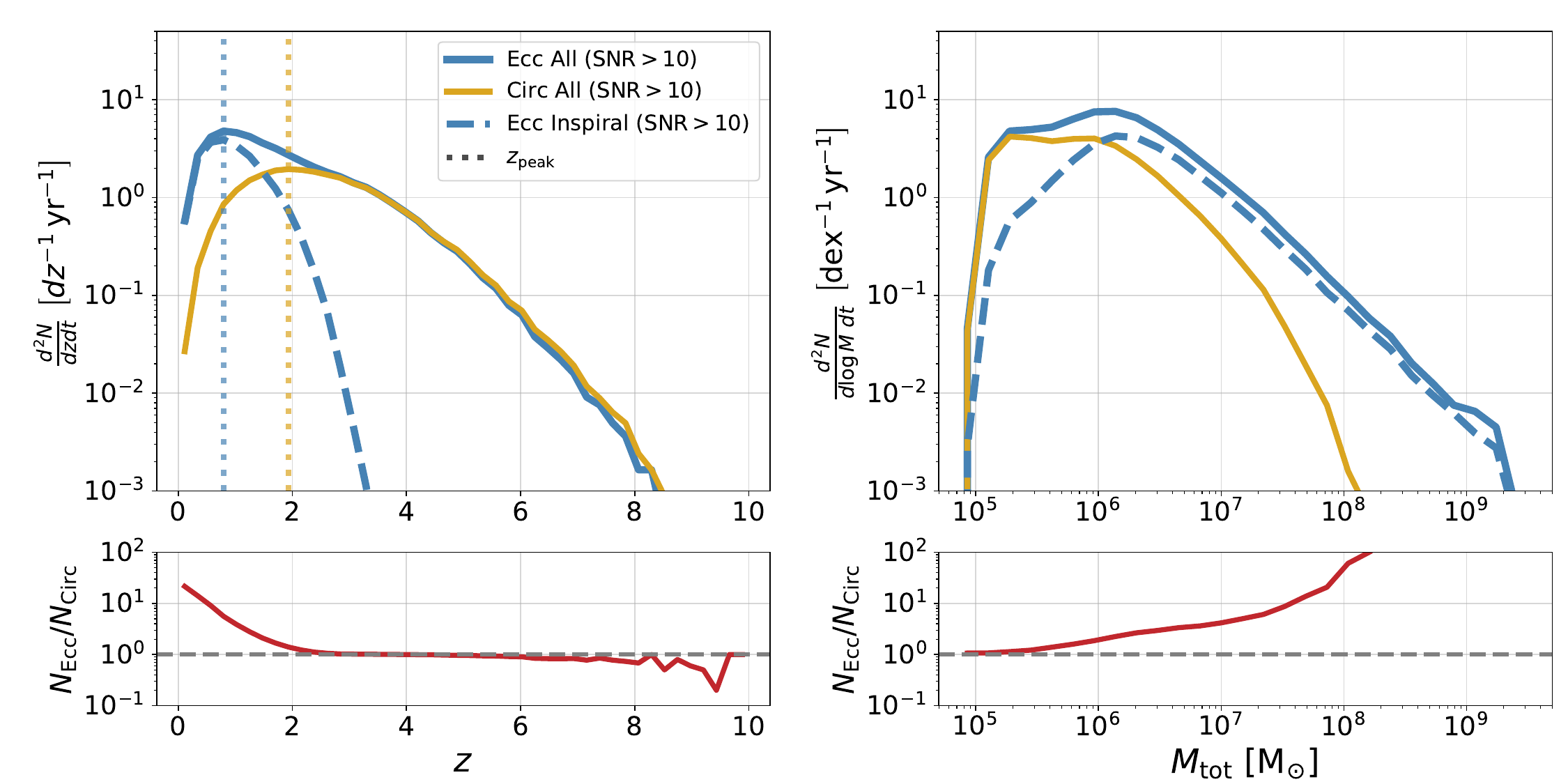}
    \caption{The LISA detection rate estimated using 10,000 realizations of MBH population. 
    $Upper:$ the number of observed binaries (with SNR $>10$) per year per unit redshift (left) / log $M_{\mathrm{BH}}$ (right). In each panel, the yellow/blue solid curves represent the results from \texttt{Circ}/\texttt{Ecc}, respectively. 
    The blue dashed curve shows the predicted inspiral sources from the \texttt{Ecc}. We do not plot the inspiral sources for the \texttt{Circ} since it has a negligible number of inspiral sources (less than $10^{-3}$/yr).
    In the upper left frame, the vertical dotted line represents the peak of redshift distribution: $z_{\mathrm{peak}}=1.9$ for \texttt{Circ} (yellow) and $z_{\mathrm{peak}}=0.80$ (blue) for \texttt{Ecc}. 
    $Bottom:$ the fraction between the number of detectable sources in \texttt{Circ} and \texttt{Ecc}. We mark the level of $N_{\mathrm{Ecc}}=N_{\mathrm{Circ}}$ with the black dashed line.
     } \label{fig:detectionRate}
\end{figure*}

\begin{figure}[!htb]
    \centering
    \includegraphics[width=1\linewidth]{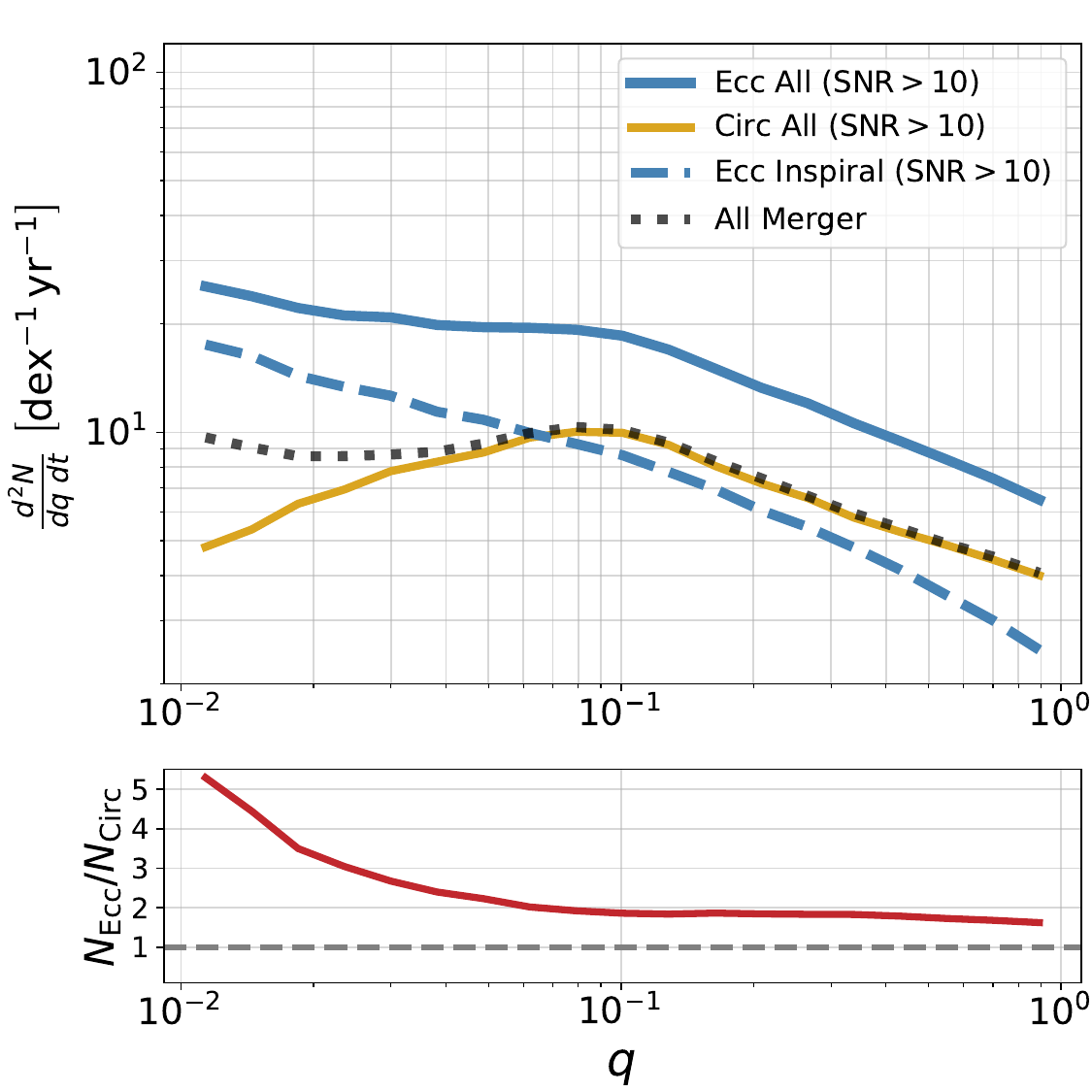}
    \caption{Same with Fig.~\ref{fig:detectionRate} but the detection rate is plotted as a function of the mass ratio $q=M_2/M_1$. The black dotted curve in the upper panel represents all the merger sources in \astrid. 
    } 
    \label{fig:DetectionRate_q}
\end{figure}

We take the waveforms calculated in Section \ref{sec:GWCalc}, and compute the strain and frequency distributions for the MBH mergers in \astrid. The results are shown in  Fig.~\ref{fig:strain_freq}, where the left panel is for the \texttt{Circ} and the right panel for the \texttt{Ecc}. 
For each MBH merger, we find the point in its time evolution and the harmonic mode that makes the greatest contribution to the SNR in the 4 years before coalescence. This is used to position the source in Fig.~\ref{fig:strain_freq}. For the \texttt{Circ}, the harmonic mode is always $n=2$.  
We color the distribution based on the redshift and the number of events, as shown by the right colorbar: the high (low) redshifts are represented by blue (red) colors, and the darkness corresponds to the regions with very few events. 
The redshift is averaged over the merger population within each pixel. 
The cut around $f\sim 10^{-2.5}$ Hz corresponds to the LISA sensitivity peak, which arises from our choice of plotting the frequency that contributes the most to the SNR. 
As we can see, the strain-frequency distributions for the \circr\ and \eccr\ are generally similar, except that the scatter in the strain range for a specific redshift is slightly larger for \eccr.

In Fig.~\ref{fig:SNR_dist}, we demonstrate the parameter space of MBH mass and redshift explored by LISA. 
The left panel shows the two-dimensional distribution of redshift and total merger mass $M_{\mathrm{tot}}$, and the right panel shows how the mass of the primary MBH $M_1$ and secondary MBH $M_2$ are distributed. 
The underlying blue color gradient represents the differential number of the detected mergers during the LISA 4-year observation for \eccr: $dN/(d\log M\, dz\,dt)
\times \mathrm{4yr}$ for the left panel and $dN/(d\log M_1\, d\log M_2\,dt)
\times \mathrm{4yr}$ for the right panel. 
The black and red contours correspond to the detected MBH merger population for \eccr\ and \circr, respectively. We can see that the detected eccentric and circular binaries largely overlap,
and most of the LISA sources are low-mass mergers with $M_{\mathrm{tot}}<10^{7}$ \Msun\ at the local Universe ($z<2$). 
An important difference between \eccr\ and \circr\ observed in Fig.~\ref{fig:SNR_dist} is that including eccentricity enables LISA to explore sources with higher mass. 
Around $z\sim 0$, the most massive mergers within detectable circular binaries have mass $\sim 10^{8}$ \Msun, while for eccentricity sources, the high-mass limit increases to $M_{\mathrm{tot}}\sim10^{9}$~\Msun (also see the right panel of Fig.~\ref{fig:detectionRate}). 
This is supported by the $M_2-M_1$ distribution plot as well: in \eccr, the covered mass spectra is increased by one order of magnitude compared to the \circr. 
\citet{Astrid_PTA_GWB} found the bulk of the GW background detected by PTAs originates from MBH binaries with masses $M_{\mathrm{tot}}=10^{9}\sim 3\times 10^{9}$~\Msun\ at low-frequency $f=4$ nHz, and from binaries within $M_{\rm tot}=10^{7.5}\sim 10^{9.3}$~\Msun at high-frequency $f=20$ nHz. Our results in Fig.~\ref{fig:SNR_dist} imply that some LISA events may also be PTA sources, although their deteactability is relatively low: based on black contours in the left panel, the probability of observing these massive mergers within the 4-year observation phase is about 0.05.  

We next make predictions for the LISA detection rate based on the prescription described in Section~\ref{sec:MergerRatePredict}. We generate 10,000 realizations of the merging MBH population, and present the averaged detection rate in Fig.~\ref{fig:detectionRate}. 
In the left (right) column, we show the number of observed sources, i.e., with the SNR $>10$, per year per unit redshift (per unit log$M_{\mathrm{BH}}$). 
In the upper panels, the solid blue and yellow curves represent the results for the \eccr\ and \circr, respectively. 
Specifically, we use the blue dashed curve to denote the contribution from the inspiral sources in the \eccr. 
We do not plot the inspiral sources for \texttt{Circ} since only a negligible number are detectable (less than $10^{-3}$/yr), which is consistent with \citet{Katz2020} (see their Table~A1). In the lower panels, we show the ratio between the number of detectable sources in \texttt{Circ} and \texttt{Ecc}. We mark  $N_{\mathrm{Ecc}}=N_{\mathrm{Circ}}$ with a black dashed line.
If we consider the detection rate per unit redshift, $d^{2}N/dz\,dt$, we observe that the \circr\ and \eccr\ peak at different redshifts: 
$z_{\mathrm{peak}}=1.9$ for \texttt{Circ} and $z_{\mathrm{peak}}=0.80$ for  the \texttt{Ecc}  (marked by the vertical yellow and blue dotted lines, respectively). 
At low redshift ($z<2$), incorporating eccentricity significantly increases the detection rate, and $z\sim0$, the ratio $N_{\rm Ecc}/N_{\rm Circ}$ is larger than 10. 
As for the detection rate across the $M_{\mathrm{BH}}$ (right column), both 
\eccr\ and \circr\ peak at the lower mass end, as expected. The plateau at $M<10^{6}$ \Msun\ and the cut at $10^{5}$ \Msun\ come from the black hole seed mass adopted in \astrid. 
In \eccr, the number of detected sources at high mass end ($M_{\mathrm{BH}}>10^{7.5}$ \Msun) is increased by more than one order of magnitude, and the part of the mass spectrum that can be explored expands from $10^{8}$ \Msun\ in the \circr\ to $10^{9}$~\Msun\ in the  \eccr.
Summed over redshift, the total LISA detection rate predicted by \astrid\ is 
$5.6$/yr for the \circr\ and $10.5$/yr for the \eccr. 
Among the events, 46\% (4.8 per year) are inspiral sources for \eccr, and only $0.3\%$ ($\sim 10^{-3}$ per year) for \circr.

An important message we can get from Fig.~\ref{fig:detectionRate} is that incorporating orbital eccentricity boosts the LISA detectability. Such an enhancement comes almost fully from the inspiral sources:
the number of detected merger sources is almost the same in the \eccr\ (5.7/yr) and in the \circr\ (5.6/yr).
In the \eccr, at low redshift ($z<0.5$) or at the high mass end $M>10^{7.5}$, the detected inspiral sources outnumber the merger sources by almost an order of magnitude. 
For the inspiral sources, as the frequency evolution during the early stages is typically slow \citep{Sesana2010}, the covered frequency range for an individual harmonic mode is very narrow, producing a smaller SNR for the inspiral sources in the \circr. 
On the other hand, in the \eccr, the harmonic modes higher than $n>2$ not only result in a wider frequency range but also extend the signal to the band where LISA has high sensitivity (see the right panel of Fig.~\ref{fig:ecc_waveform}). This makes the inspiral sources more likely to be detected when the eccentricity is taken into account. 

A caveat is that at $z>2$ or $M_{\mathrm{BH}}<10^{6}$ \Msun, the number of detectable inspiral sources drops quickly. This is because at high redshift, or the low mass end, the strain is too low to produce a detectable signal even if the high harmonic mode is close to the LISA high-sensitivity band.
Moreover, for the eccentricity binaries, their peak strain is lower than their circular counterpart, which can be seen from Fig.~\ref{fig:ecc_waveform} (middle and right panels). 
This explains why at high redshift ($z>6$), \circr\ predicts a slightly larger detection rate than the \eccr. 
The mass peak for the inspiral sources at $M_{\mathrm{tot}}\sim10^{6}$~\Msun\ shifts the entire distribution of detectable sources in the \eccr\ to slightly higher mass. 
In the \circr, where almost all the sources are merger sources, the mass peak is around the black hole seed mass $M_{\mathrm{tot}}\sim10^{5}$ \Msun.

The minor increase in the number of detected merger sources in the \eccr\ (from $5.6$/yr in \circr\ to $5.7$/yr) is from massive mergers. With the circular orbit assumption, no mergers with $M_{\mathrm{tot}}>10^{8}$ \Msun\ can be detected, while in the \eccr, there are still some sources with $M_{\mathrm{BH}}\gtrsim 10^{9}$ \Msun\ observed through coalescence in the 4-year observation. This corresponds to the gap between the blue solid curve and the blue dashed curve.
Binaries with such a large mass typically merge at low frequency before they are observed in the LISA band. When we include the eccentricity, the higher harmonic modes increase the emission at higher frequencies. This is the case plotted in the right panel of Fig.~\ref{fig:ecc_waveform}.

In Fig.~\ref{fig:DetectionRate_q}, we present the detection rate as a function of the MBH mass ratio $q=M_2/M_1$. In the upper panel, besides all the detected sources in \eccr\ (blue solid curve), \circr\ (yellow solid curve), and the inspiral sources in the \eccr\ (blue dashed curve), we also plot the total merger rate in \astrid\ (black dotted curve). 
Most of the LISA sources have a low mass ratio: $q<0.1$. This is because there are more low-$q$ mergers in \astrid\ but not because LISA is more sensitive to them. Actually, LISA can observe almost all the mergers with $q\geq 0.1$, which is shown by the overlap between the yellow curve and black curve, while a large fraction of low-$q$ mergers is not detectable. 
Note that the mass ratio has a dependence on the merger mass $M_{\mathrm{BH}}$: low-$q$ mergers are dominated by massive mergers, since the secondary MBH mass $M_2$ is still massive than the adopted black hole seed mass $10^{4.5}$ \Msun. This dependence explains why the detection rate for the \circr\ drops at low-$q$ end, and the enhancement of the detection rate in \eccr\ at $q<0.1$ is consistent with the right panel of Fig.~\ref{fig:detectionRate}.

\begin{figure*} [htp]
    \centering
    \includegraphics[width=1\linewidth]{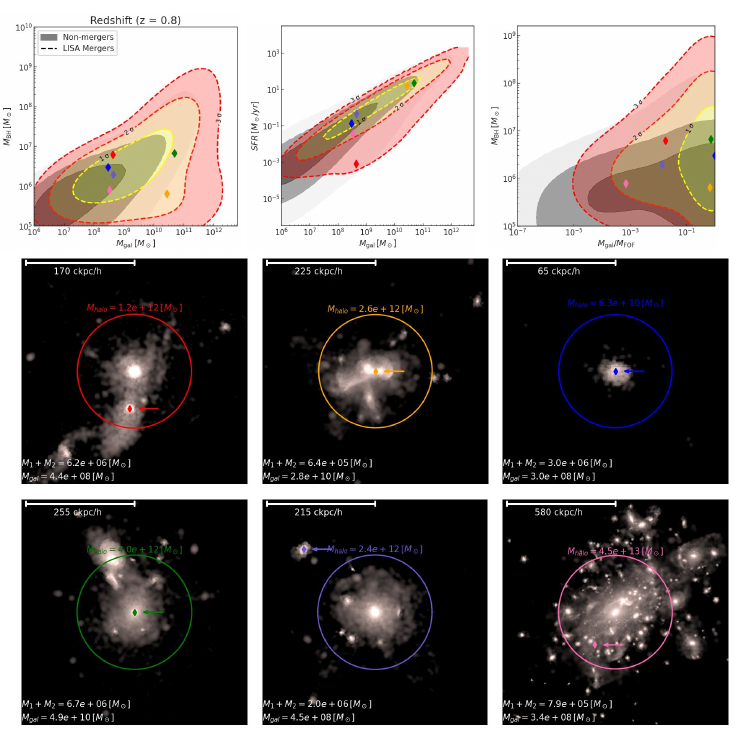}

    \caption{
    The galaxies hosting detectable sources at $z=0.8$.
    The top contour plots compare the 
    properties of host galaxies of detectable LISA sources which merger within $0.8 <z<0.9$ (red contours) and the galaxies hosting no mergers (grey contours).
    Each contour set plots the $1\sigma/2\sigma/3\sigma$ regions, enclosing 68\%/95\%/99.7\% of the population. \textit{Top Left}: The MBH mass versus galaxy stellar mass ($M_{\mathrm{BH}}-M_{\mathrm{gal}}$).
    \textit{Top Middle}: The galaxy star-formation rate versus galaxy stellar mass ($\mathrm{SFR}-M_{\mathrm{gal}}$). 
    Note that we remove the galaxies with zero star formation rate to plot this distribution. 
    \textit{Top Right}: The MBH mass versus the stellar mass ratio between the host galaxy and host Friends-of-Friends (FoF) halo ($M_{\mathrm{BH}} - M_{\mathrm{gal}}/M_{\mathrm{FoF}}$).
    For the galaxies hosting detectable sources, we plot the mass of the remnant MBH as $M_{\mathrm{BH}}$; and for the non-merger galaxies, we plot the mass of the central MBH. 
    We randomly select a sample consisting of 6 detectable sources, which are marked by the diamonds in the contour plots. 
    In the second and third rows, we plot the stellar density field for the FoF halo for the selected mergers.
    The location of the remnant MBH is marked by the arrows with corresponding colors. 
    In each frame, the circle represents the virial radius of the halo. 
    }
    
    \label{fig:Contours1}
\end{figure*}

\section{Host Galaxy and Environment}
\label{sec:hostgal}

\begin{figure*} [htp]
    \centering
    \includegraphics[width=1\linewidth]{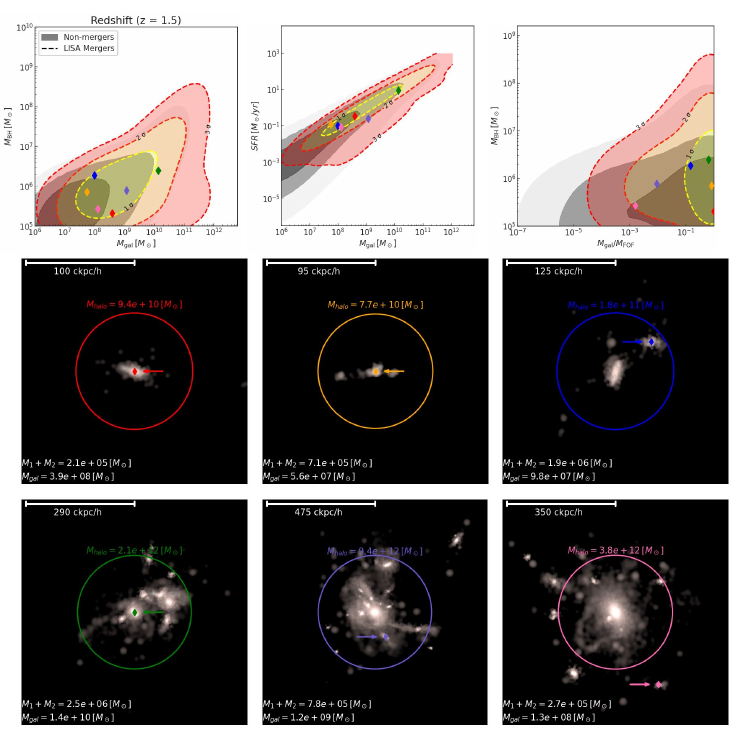}
    \caption{
    Same with Fig.~\ref{fig:Contours1}, but for the host galaxies of detectable mergers occur within $1.5<z<1.6$.
    }
\label{fig:Contours2}
\end{figure*}

In this section, we focus on the merging populations at two redshifts: $z=1.5$ and $z=0.8$ to investigate the correlation between the mergers and their local environments.

\subsection{Host Galaxies of Detected Sources}

Identifying host galaxies or AGN activities of gravitational wave sources is key to the multi-messenger study of MBHs and has the potential for constraining cosmology.
Future detectors like LISA will be able to localize the GW sources to the square-degree level, much better than the current ground-based detectors.
However, there are still up to a few hundred candidate galaxies in the LISA source field \citep{Lops2023}. 
Being able to distinguish the merger hosts greatly facilitates the EM-followup of LISA events.
Recently, \citet{Bardati2024a} and \citet{Bardati2024b} investigated the morphological signatures of MBH merger hosts using cosmological simulation Romulus \citep{Romulus_paper} and found persistent morphological changes in high-mass major mergers.
\citet{Izquierdo-Villalba2023} investigated the signatures of LISA source hosts using SAM and found it difficult to identify the merger hosts solely from the galaxy properties.
In this section, we look for distinctive features of MBH merger hosts from the wide range of galaxy population from ASTRID.

In Fig.~\ref{fig:Contours1} and~\ref{fig:Contours2} we present the properties of the galaxies hosting LISA sources, and compare them to the galaxy population hosting no mergers at $z = 0.8$ and $z = 1.5$, respectively. 
In the top row of both figures, the red contours represent the galaxy hosting the MBH mergers detectable by the LISA, i.e., SNR $>10$, whose boundaries are highlighted by the dashed lines; and the contours with a grey color scheme correspond to the galaxy population not hosting mergers at that redshift. 
For each contour set, we plot the  $1\sigma/2\sigma/3\sigma$ regions, enclosing 68\%/95\%/99.7\% of the population.   
From left to right, we show the two-dimensional distribution of the MBH mass versus galaxy stellar mass ($M_{\mathrm{BH}}-M_{\mathrm{gal}}$, top left), galaxy star-formation rate versus galaxy stellar mass ($\mathrm{SFR}-M_{\mathrm{gal}}$, top middle), and MBH mass versus the stellar mass ratio between the host galaxy and host Friends-of-Friends (FoF) halo ($M_{\mathrm{BH}} - M_{\mathrm{gal}}/M_{\mathrm{FoF}}$, top right). 
For the galaxies hosting detectable sources, we plot the mass of the remnant MBH as $M_{\mathrm{BH}}$; and for the non-merger galaxies, we plot the mass of the central MBH. 
At both redshifts, we randomly select a sample consisting of six detectable sources, whose positions among the two-dimensional contours are marked by diamonds. In the middle and bottom rows, we plot the stellar density field for the FoF halo for the selected mergers.
The location of the remnant MBH is marked by diamonds with the color used for the same object in the contour panel, and the circle represents the virial radius of the halo.

At both $z=0.8$ and $z=1.5$, the galaxies hosting detectable merging sources lie on the same $M_{\mathrm{BH}}-M_{\mathrm{gal}}$ correlation as the non-merger galaxies, but with differences in how they populate this and the rest of the parameter space.
Compared to the non-merger galaxy population, merger hosts have larger galaxy mass on average, with the center of the distribution lying at $10^{9}$ \Msun\ at both redshifts plotted. 
This is two orders of magnitude beyond the mean of the non-merger distribution, the latter of which does depend on the lower mass limit for inclusion into the sample.  
A significant fraction of the merger hosts are also very massive galaxies with large BHs, as the contour enclosing 99.7\% of the population extending to $M_{\mathrm{gal}}\sim 10^{12}$ \Msun\ and $M_{\mathrm{BH}}\sim 10^{9}$~\Msun. 
The contours also extend to lower masses, showing that a wide variety of galaxies can be hosts. For example, many detectable mergers of small MBHs take place in galaxies with masses as small as $10^7$~\Msun\ and below.
Our mass range of the merger hosts is similar to those predicted by recent work using SAMs \citep[e.g.][]{Izquierdo-Villalba2023}.

The ${\rm SFR}-M_{\mathrm{gal}}$ distributions are similar among the host galaxies hosting LISA mergers and those that are not. Note however that LISA mergers host galaxies occupy the more massive end of the population. 
LISA merger host galaxies also have SFRs ranging from $0.07\sim5.56$~\Msun/yr, (almost none have a SFR smaller than $10^{-3}$~\Msun/yr, even for the low-mass end). Some merger host galaxies have a high SFR of 10-1000 \Msun/yr, and if we examine two of these (the orange and green symbols in Fig.~\ref{fig:Contours1}), we can see that while both lie at the centers of massive galaxies, one (shown with orange symbok) is a disturbed ongoing galaxy merger, whereas the other (green) is in a much more uniform (yet somewhat disturbed) galaxy.

The top right panels of Fig.~\ref{fig:Contours1} and \ref{fig:Contours2} show that most of the host galaxies of LISA mergers are the central galaxies within a halo.  For example, 68\% of all merger hosts have a stellar mass greater than $\sim 10 \%$ of that of the FOF halo they reside in.
However, a significant fraction of the merger hosts ($\sim$ 20-30\%) does lie in the satellite galaxies.
For example, among the six examples we have selected and plotted, only half of them are located in the centers of their parent halos, and one of them at each redshift is even hosted by galaxies outside the virial radius (the indigo diamond in Fig.~\ref{fig:Contours1} and the pink diamond in Fig.~\ref{fig:Contours2}).
The high probability of merger hosts in isolated systems (i.e., with $M_{\mathrm{gal}}/M_{\rm FOF}$ close to 1) may facilitate the EM identification of the LISA sources among the $>100$ candidate galaxies in the LISA source field \citep[e.g.][]{Lops2023}.

Our results reveal that LISA will observe MBH mergers in various galaxy environments, covering a wide parameter space of stellar mass and SFR, and in central or satellite galaxies. 
This is significantly different from the target sources for PTAs. 
As found in \citet{Zhou2025_PTACW}, the gravitational continuous (CW) wave sources that have high detection probability for PTAs are all observed in the central galaxies located at the center of galaxy clusters. All of these PTA CW host galaxies have stellar mass over $10^{12}$ \Msun, and SFR$\gtrsim200$ \Msun/yr.  
Hence, LISA detections will provide complementary information to the MBH mergers that occur in lower-mass galaxies. 

\begin{figure*}
    \centering
    \includegraphics[width=1\linewidth]{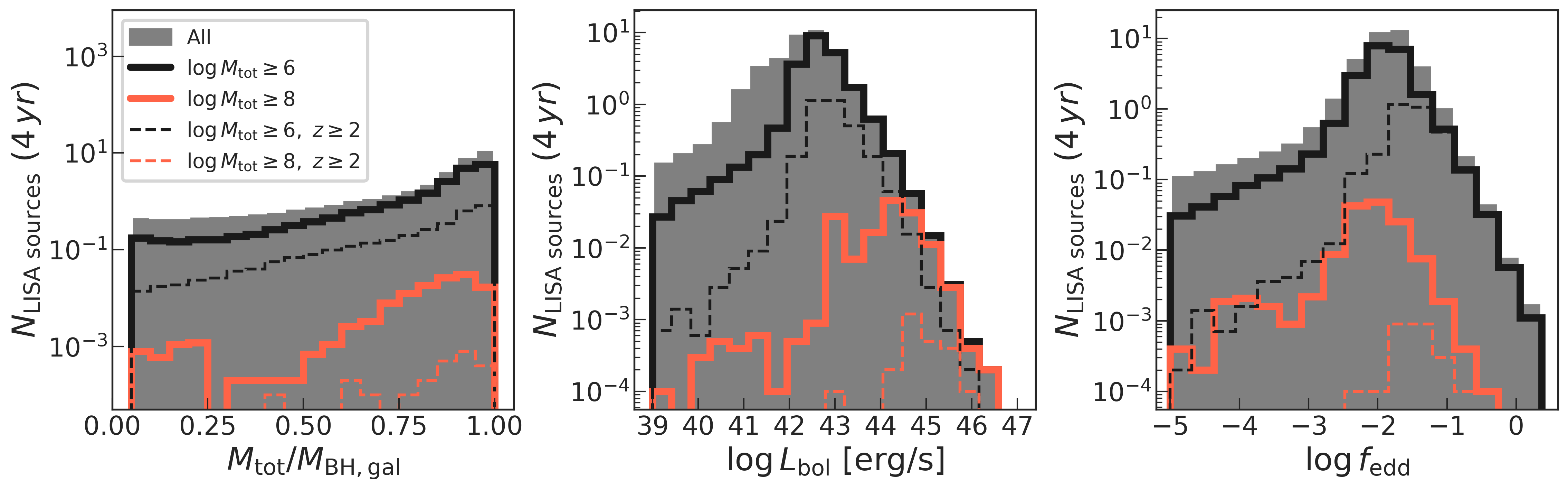}
    \caption{
    The MBH population in the detectable mergers (with SNR $>10$) in the \eccr. \textit{Left}: Ratio of total merger mass ($M_{1} + M_{2}$) to the total black hole mass in the host galaxies. \textit{Middle}: MBH Bolometric luminosity $L_{\mathrm{bol}}$. \textit{Right}: Black hole Eddington ratio $f_{\mathrm{edd}}$.
    In each frame, the grey histogram represents the distribution of all the mergers across the redshift and mass range. 
    The solid black(red) curves show the MBHs involving mergers with $M_{\mathrm{tot}}>10^{6}$~\Msun($M_{\mathrm{tot}}>10^{8}$~\Msun). 
    The dashed histograms correspond to the mergers occurring at high redshift ($z\geq 2$) with $M_{\mathrm{tot}}>10^{6}$ \Msun\ (black dashed) or $M_{\mathrm{tot}}>10^{8}$ \Msun\ (red dashed).
    }
\label{fig:Env_1D}
\end{figure*}

\subsection{MBH Population Detected by LISA}

In this section, we investigate the MBH population involved in the mergers detected by LISA. In Fig.~\ref{fig:Env_1D}, we show the PDF for the ratio of the merging MBH mass $M_{\mathrm{tot}}=M_{\mathrm{1}}+M_2$ to the total MBH mass in the host galaxy (left), 
the bolometric luminosity $L_{\mathrm{bol}}$ (middle), and
the Eddington ratio $f_{\mathrm{edd}}$ (right). 
In each frame,
besides showing all the detectable mergers in the \eccr\ across the whole redshift range using the grey histogram, we also plot the distribution of MBHs involving mergers with $M_{\mathrm{tot}}>10^{6}$~\Msun\ (black solid curve) and $M_{\mathrm{tot}}>10^{8}$~\Msun\ (red solid curve). 
The dashed histograms correspond to the mergers occurring at high redshift ($z\geq2$) with $M_{\mathrm{tot}}>10^{6}$~\Msun\ (black) or $M_{\mathrm{tot}}>10^{8}$~\Msun\ (red).
$M_{\mathrm{tot}}>10^{6}$~\Msun\ is where the detection rate peaks in the right panel of Fig.~\ref{fig:detectionRate}, so the black solid curve represents most of the mergers, and the gap between the black solid curve and the grey histogram is the contribution from the low-mass mergers.

From the distribution of $M_{\mathrm{tot}}/M_{\mathrm{BH,gal}}$, we can see that in most host galaxies (85\%) merging binaries constitute over half of the total black hole mass, which means the remnant MBHs are typically the galaxy's central black hole. 
The distribution is similar among the whole population and the high-redshift mergers, while for mergers at the high-mass end ($M_{\mathrm{tot}}>10^{8}$~\Msun), 
a larger fraction of the remnant MBHs (96\%) dominates the galaxy black hole mass with $M_{\mathrm{tot}}/M_{\mathrm{BH,gal}} > 0.5$.

We then investigate the accretion of the merging MBHs.
For all the detectable mergers in \eccr\ and the population with $M_{\mathrm{tot}}>10^{6}$~\Msun, the 
$L_{\mathrm{bol}}$ distribution both peak at $\log(L_{\mathrm{bol}})=42.6\,\mathrm{erg/s}$. 70\% of total mergers have $\log(L_{\mathrm{bol}})>42$ erg/s. 
As expected, massive mergers in general have higher luminosity: $L_{\mathrm{bol}}$ peaks at $10^{44.3}$ erg/s for mergers with $M_{\mathrm{tot}}>10^{8}$~\Msun. 
Only a small fraction of mergers have relatively high luminosity: 
less than $1\%$ detected mergers have luminosity larger than $10^{44}$ erg/s, and 66\% of them are from merger with 
$M_{\mathrm{tot}}>10^{8}$~\Msun. Recently, \citet{Paez2023} also investigated the AGN signatures of LISA sources, but at a higher redshift of $z\sim 3.5$, and found $4\sim 20\%$ of sources to be detectable through AGN signatures.
Our results are in general agreements despite the different source redshift distributions.

For the Eddington ratio $f_{\rm edd}$, there is only a minor distinction between the population with different mass cuts, while it presents evolution across the redshift: 
for both the merger with $M_{\rm tot} \geq 10^{6}$~\Msun\ and $M_{\rm tot} \geq 10^{8}$~\Msun\ over the whole redshift range, $f_{\rm edd}$ peaks at $0.01$; while for $z\geq2$, they have higher Eddington ratio, both peaking at $0.02$. 
Note that $f_{\rm edd}$ can be larger than 1 since we adopt a super-Eddington accretion.

\section{Discussion}
\label{sec:discussion}

\subsection{Binary Delay Timescale}

In this work, we do not apply any binary delay model to account for the evolution timescale from a separation of $1$ kpc to the GW regime.
This choice is motivated by the fact that \astrid\ already includes the effects of DF, which in previous works have been shown to dominate delays in MBH binary dynamics (prior to the GW-driven phase). 
In particular, \citet{Li2022_tng50} evolved unbound MBH pairs in the TNG50-3 cosmological simulation from kpc scales to coalescence using a detailed postprocessing formalism that incorporates DF, loss-cone scattering, viscous drag from a circumbinary disk, and GW emission.
Their results showed that the DF-dominated phase accounts for the largest fraction of the total binary hardening timescale.

Several previous studies incorporated binary delay based on postprocessing the simulation binaries with analytical prescriptions to predict MBH merger rates and associated  LISA event rates.
\citet{Katz2020} post-processed the MBH merger population from the Illustris cosmological simulation using the prescriptions of \citet{Kelley2017b} and \citet{Dosopoulou2017}.
They found that considering the delay timescale only reduced the detection rate by at most a factor of two.
Similarly, \citet{Salcido2016} concluded that merger rates are relatively insensitive to the specific adopted hardening model, and the choice of MBH seed mass has a much stronger influence. 
\citet{Volonteri2020} directly compared the binary population subject only to DF with those incorporating additional hardening mechanisms, including stellar scattering and gas torques.
Those authors also found that these additional effects have a minor impact on the predicted detection rate, significantly smaller than that of DF delay (see their Fig.~8).
Together, these findings suggest that incorporating a binary delay model would only modestly alter the predicted detection rate shown in Fig.~\ref{fig:detectionRate}.

Self-consistently tracking the binary dynamics on small scales requires simulations with higher resolution. The MAGICS simulation suite \citep{Chen2024, Zhou2025, Mukherjee2025} resimulates 15 merging systems extracted from \astrid\ with the resolution up to 62.5~\Msun, and followed the evolution from $10$ kpc separation into hard binary phase ($\sim 10^{-3}$ pc). Although by now MAGICS has only investigated a small sample out of the merger population in \astrid, upcoming works will include merging systems covering a larger parameter space. 
Simulations like MAGICS that directly resolve the binary formation and evolution will continue to provide valuable insights to evaluate and calibrate the results from the large-volume simulations.

\subsection{The Effect of Orbital Eccentricity}

In this study, we initialize the binary eccentricity for LISA-band using the value measured on the scales of $\sim$1 kpc in \astrid. We do not evolve the eccentricity through (i) the stellar scattering, or (ii) the early GW-driven circularization before detection.

For the pre-observation GW circularization, we expect that it has minimal influence on our LISA predictions. 
Circularization during the inspiral phase is typically slow until the final few orbits before the coalescence \citep{Peters1964, Wagg2022}. 
Fig.~\ref{fig:detectionRate} shows that the increased LISA detectability in the \eccr\ arises nearly entirely from inspiral sources.
Those systems experience only the early-phase circularization prior to observation, which contributes little to eccentricity evolution.

Stellar scattering during the MBH binary hardening phase can lead to an increase in orbital eccentricity before the GW-dominated regime \citep{Sesana2010, Sesana2008}.
MAGICS suites give a self-consistent description of the orbital eccentricity evolution during the hardening phase \citep{Zhou2025, Mukherjee2025}.
It observes the slight enhancement of eccentricity for seed MBH mergers whose initial eccentricity $e_0 > 0.6$, while for those with $e_0<0.6$, it remains nearly the same. 
Neglecting this evolution therefore makes our adopted eccentricities serve as a lower limit of the realistic values. 
Consequently, our results effectively provide a lower bound on the impact of eccentricity.

\section{Conclusion}
\label{sec:conclusion}

In this work, we estimate the GW signal emitted by MBH mergers down to $z=0$ in the cosmological simulation \astrid, and make predictions for the LISA detection rate. 
Using the recorded trajectories of MBHs in \astrid, we are able to measure the orbital eccentricity $e_0$ of the binaries at scales of $\sim 1\ \rm{kpc}$.
We find that the MBH mergers in general have high eccentricity: $e_0$ peaks at $\gtrsim 0.8$ across the whole redshift range. It evolves slightly to smaller values at low redshifts: from $0.88$ at $z>3$ to $0.80$ at $z<0.5$, indicating that a larger fraction of binaries at low redshifts merge with circular orbits. This can be explained by the shorter hardening timescales for binaries with high eccentricity \citep{Chen2022}. 
To give a better description of the $e_0$ distribution,
we provide a fitted $e_0$ PDF function (Equation~\ref{equ:fitted_ecc_pdf}), which includes the redshift evolution.

To investigate the influence of eccentricity on GW emission in the LISA band ($10^{-5}\,{\rm Hz}<f<1\, \rm Hz$) 
we generate two sets of signals:  a \circr\ that assumes circular orbits for all the sources, and a \eccr\ that includes the effect of eccentricity in the calculation of GW waveforms. 
Given the  black hole seed masses in \astrid\ of $M_{\mathrm{BH,seed}}\sim 5\times10^{4}$ \Msun\ and a SNR threshold of 10, we find that LISA will probe MBH mergers across a wide parameter space, covering mergers with $M_{\mathrm{BH}}\lesssim 10^{9}$ \Msun\ and $z\lesssim 8$.

We implement a Monte Carlo analysis technique
to estimate the LISA detection rate.
This enables us to generate multiple realizations of the merger population detectable by LISA within the 4-year observation, and avoids the assumption of a specific observation time for each MBH binary. 
We can therefore consider the signals emitted from different binary evolution phases, i.e., from both inspiral sources and merging sources. 
We generate 10,000 realizations of the merging MBH populations that would be detected by a 4-year LISA observation. 
Our prediction is averaged over these realizations: the total LISA detection rate predicted by \astrid\ is $5.6$/yr for the \circr\ and $10.5$/yr for the \eccr. 
Among them, 46\% (4.8 events per year) are inspiral sources for \eccr, and only $0.3\%$ ($\sim 10^{-4}$ event per year) are for \circr. Our predicted number of detections is noticeably higher than those found in previous studies based on large-volume hydrodynamical simulations such as EAGLE \citep{Salcido2016} and Illustris \citep{Katz2020}, which reported rates of $\sim 0.5$--$2$ per year. This also places our results closer to those from semi-analytical models, which have generally predicted higher detection rates ( $\sim 8$--$25$ per year) \citep{Bonetti2019, Berti2016, Klein2016, Sesana2011, Arun2009, Izquierdo-Villalba2024}, suggesting a convergence between these two modeling approaches.

Including the eccentricity will expand the detected mass range by one order of magnitude: from $M_{\mathrm{BH}}\sim10^{8}$~\Msun\ in the circular scenario to $\sim 10^{9}$~\Msun, and shift the peaks of redshift of LISA sources to a lower value: $z_{\rm peak}=0.8$ in \eccr\ compared to the $z_{\rm peak}=1.9$ in \circr. 
At low redshift ($z<0.5$) and high mass end ($M_{\mathrm{tot}}>10^{7.5}$ \Msun), the \eccr\ predicts a detection rate one order of magnitude higher than \circr. 
The improvement in detectability comes mainly from the inspiral sources, which coalesce after the 4-year LISA observation phase. 
The residual eccentricity in the inspiral phase produces waveforms covering a wide frequency range and GW signals better matching the LISA high-sensitivity band. 
Our results underscore the importance of incorporating the orbital eccentricity in the predictions for upcoming GW detection, which requires accurately modeling the dynamics of MBH binaries.

We also investigate the host galaxies of detectable mergers. 
Unlike the PTA CW sources, LISA detectable mergers will be observed to lie in various galaxy environments, covering a wide range of mass ($10^{7} - 10^{11}$ \Msun), and SFR ($10^{-3}-10^{3}$ \Msun/yr).
Notably, a significant fraction ($\sim20-30\%$) of the detectable sources occur in satellite galaxies, sometimes even outside the virial radius.  
Most of the remnant MBHs are the most massive MBH of the host galaxy.
For all the detectable mergers in \eccr, the distribution of bolometric luminosity peaks at $L_{\rm bol}=10^{42.6}\ \rm erg/s$. Massive merger with $M_{\mathrm{tot}}>10^{8}$~\Msun\ in general have larger luminosity, peaking at $L_{\rm bol}=10^{44.3}\ \rm erg/s$. 
Only 1\% of detectable mergers have luminosity larger than $10^{44}\ \rm erg/s$, 66\% of which are mergers with $M_{\mathrm{tot}}>10^{8}$~\Msun.

\section*{Acknowledgements}
YZ acknowledges helpful discussion with Lei Hu. 
BYW thanks Yili Han and Handan Zhang for their helpful advice on the visualization. TDM acknowledges funding from NASA ATP 80NSSC20K0519, NSF PHY-2020295, NASA ATP NNX17AK56G, and NASA ATP 80NSSC18K101, NASA Theory grant 80NSSC22K072.
NC acknowledges support from the Schmidt Futures Fund. 
YN acknowledges support from the ITC Postdoctoral Fellowship.
SB acknowledges funding from NASA ATP 80NSSC22K1897.
\astrid~was run on the Frontera facility at the Texas Advanced Computing Center.

\bibliography{main}{}
\bibliographystyle{aasjournal}
\end{CJK*}
\end{document}